# Electronic structure of $LaFe_{1-x}Co_xAsO$ from first principle calculations

Haiming Li[1], Jiong Li[1], Shuo Zhang[1,2], Wangsheng Chu[1], Dongliang Chen[1] and Ziyu Wu[1,2,3,*]

[1]*Beijing Synchrotron Radiation Facility, Institute of High Energy Physics, Chinese Academy of Sciences, Beijing 100049, P. R. China*

[2]*National Synchrotron Radiation Lab, University of Science and Technology of China, Hefei 230026, P. R. China*

[3]*Theoretical Physics Center for Science Facilities (TPCSF), Chinese Academy of Sciences, Beijing 100049, P. R. China*

Based on the first-principles calculations, we have investigated the geometry, binding properties, density of states and band structures of the novel superconductor $LaFe_{1-x}Co_xAsO$ and its parent compounds with the ZrCuSiAs structure. We demonstrate that La–O bond and TM-As (TM=Fe or Co) bond are both strongly covalent, while the LaO and TMAs layers have an almost ionic interaction through the Bader charge analysis. Partial substitution of iron with cobalt modify the Fermi level from a steep edge to a flat slope, which explains why in this system Co doping suppresses the spin density wave (SDW) transition.

PACS numbers: 74.25.Jb, 74.70.-b, 71.20.-b

---

[*] Electronic address: wuzy@ihep.ac.cn

# Introduction

Following the discovery of high-$T_c$ copper oxides[1] and $MgB_2$[2], the appearance of superconductivity in quaternary pnictide-oxides triggered new researches and stimulated the entire scientific community. Of this new class of compounds, iron- and nickel-based layered compounds LaFePO[3] and LaNiPO[4] were the first systems that showed superconductivity although with a low transition temperature $T_c$: 5 K and 3 K, respectively. In the last months, the superconductivity transition temperature raised first 26 K in the LaFeAsO by a partial substitution of O with F atoms ($LaFeAsO_{1-x}F_x$)[5], and 43 K at high pressure[6]. The progress encouraged researchers to look for superconductivity in other FeAs-based materials and other families of FeAs quaternary pnictide-oxides with Ce, Pr, Nd, Sm, Gd replacing La were identified with superconducting temperature higher than 50 K[7-12]. The quaternary pnictide-oxides LaFeAsO belong to a tetragonal family with the ZrCuSiAs type structure and the space group of P4/nmm[13]. They consist of alternative stacking of FeAs tetrahedral layers and LaO tetrahedral layers along the c-axis. These two layers are supposed to be positively and negatively charged, respectively so that the LaO layer mainly acts as a donor and superconducting pairing is supposed to occur in the FeAs layers. The parent materials without doping undergo a weakly first order structural phase transition from tetragonal (P4/nmm) to orthorhombic (Cmma), and then followed by antiferromagnetic spin density wave (SDW) transition at about 140 K[14]. More recently, it has been reported that other free oxygen FeAs-based compounds: $Ba_{1-x}K_xFe_2As_2$[15-17], $Sr_{1-x}K_xFe_2As_2$[18], $Eu_{1-x}K_xFe_2As_2$[19], $Ca_{1-x}Na_xFe_2As_2$[20], $Li_{0.6}FeAs$[21] may exhibit a similar superconducting behavior as the $LaFeAsO_{1-x}F_x$, with $T_c$ of 38 K, 38 K, 32 K, 20 K and 18 K, respectively. The similar SDW instablity is also observed in these systems. The maximum $T_c$ achieved up today is 56 K by with the partial substitution of Gd by Th in the GdFeAsO[22]. At the moment it is not possible to predict if $T_c$ in FeAs based compounds will continue to grow as rapidly as happened in cuprate superconductors, nevertheless this unexpected discovery will be certainly useful to better understand the physics of high-$T_c$ superconductor materials. The suppression of SDW transition is actually believed to be an important prerequisite for the appearance of superconductivity in these doped FeAs-based materials[8, 14, 23, 24].

In general, superconductors are doped in the donor layers by nonmagnetic atom, as in the case of $La_{2-x}Ba_xCuO_4$[1] and $Na_xCoO_2 \cdot yH_2O$[25], etc.. Indeed, it was well established that doping by magnetic atoms in a conducting layer generally destroys Cooper pairs and induces large distortion in the layer. However, a new type of FeAs based superconductor: the $LaFe_{1-x}Co_xAsO$ with $T_c \sim 10$ K is just the result of a real doping by a magnetic atom (cobalt) in the superconducting-active FeAs layer, and the doping system suppress the SDW transition[26, 27]. In addition, Co-doped $SrFe_2As_2$[28] and $BaFe_2As_2$[29] show also superconductivity at $T_c \sim 20$ K and 22 K, respectively. The evidence of Co-doping inducing superconductivity has challenged our knowledge between superconductivity and magnetic interactions. Additional accurate investigations are then mandatory.

Theoretical investigations in the frames of the density functional theory successfully reproduced the slope of the density of states (DOS) near the Fermi level, and gave many reasonable explanations on the interaction of FeAs-based superconductors[24, 30-35]. In this study, we focus through first principle calculations, to both geometry and electronic structure near the Fermi level of the new iron-pnictide system $LaFe_{1-x}Co_xAsO$, and explain why in this system Co doping suppresses the spin density wave (SDW) transition.

## Calculating method

The present calculations have been performed using the first-principles plane-wave Vienna *ab initio* simulation package (VASP)[36, 37] while the exchange-correlation is described by the Perdew-Burke-Ernzerhof general gradient approximations (GGA)[38]. The projector augmented wave (PAW) method in its implementation of Kresse and Joubert was used to describe the electron-ion interaction[39, 40].

The La ($5s^25p^65d^16s^2$), Fe ($3d^64s^2$), Co ($3d^74s^2$), As ($3s^23p^3$), O ($2s^22p^4$) are treated as valence states. To ensure an enough convergence, the energy cutoff was chosen to be 600 eV, while the Brillouin zone was sampled with a mesh of 16×16×8 k points generated by the Monkhorst–Pack[41] scheme for the pure LaFeAsO and LaCoAsO. Doping was modeled with a supercell of the parent material with Co atoms, with well converged grid k points in calculations. A first-order Methfessel–Paxton method with σ=0.2 eV has been considered for the relaxation[42]. The crystal cell and the internal parameters were optimized using the conjugate gradient method until the total forces on each ion was less than 0.02 eV/Å. The density of states (DOS) calculations were performed using the tetrahedron method with the Blöchl corrections[43].

## Results and discussion
**I Pure LaFeAsO and LaCoAsO systems**

**A. Crystal structure**

LaFeAsO and LaCoAsO crystallize in a tetragonal structure with the space group P4/nmm. Corresponding Wyckoff positions of the space group for different atoms are La(2c) (0.25, 0.25, $z_{La}$), Fe or Co(2b) (0.75, 0.25, 0.5), As(2c) (0.25, 0.25, $z_{As}$), and O(2a) (0.75, 0.25, 0), where $z_{La}$ and $z_{As}$ are both internal coordinates. Partial replacement of Fe with Co atoms in the LaFeAsO structure generates the LaFe$_{1-x}$Co$_x$AsO whose crystal structure is displayed in Fig. 1.

Table 1 reports the calculated lattice constants and the internal coordinates of both LaFeAsO and LaCoAsO together with available experimental data. The experimental determinations are consistent and in good agreement with our calculations. The existing differences among experimental data and calculations may be addressed to the poor description of the exchange-correlation interaction by both LDA and GGA approaches in density functional theory. Both LaFeAsO and LaCoAsO systems are layered structure with alternating stack of LaO and FeAs (CoAs) layers. FeAs (CoAs) layers are conducting layers formed by a square lattice sheet of Fe (Co) ions coordinated by As above and below the plane to form face sharing FeAs$_4$ (CoAs$_4$) tetrahedra[24]. The Fe (Co) atoms coordinate tetrahedrally with four As atoms with a bond length of 2.34 Å, forming a distorted tetrahedra with two different As-Fe-As (As-Co-As) angles of 118.86° and 104.99° (119.58° and 104.67°), in agreement with results of Refs. [5, 14, 44]. Every Fe (Co) has also four neighboring Fe (Co) atoms within the same layer with a bond length of about 2.85 Å (2.86 Å).

**B. Electronic structure**

The LaTMPnO (Pn=P, As) systems crystallize in quasi two-dimensional structure and the LaO and TMPn layers interact through an ionic interaction. For a deeper understanding of the framework of LaFeAsO and LaCoAsO systems, we performed also a Bader analysis of the charge density[45, 46]. Table 2 compares the charges of each atom using the Bader analysis with the pure ionic picture. It addresses that the large charge transfer of both La-O bonds and TM-As bonds, which is quite different from the ionic description, are strong chemical bonding. Moreover, the charge transfer between LaO and TMAs layers is considerably smaller, which implies an ionic bonding between layers. Similar conclusions have also addressed in both LaFePO[47] and LaNiPO[31].

The density of states of LaFeAsO and LaCoAsO compound was calculated. Fig. 2 shows that the DOS lineshape of these compounds is quite similar and in the case of LaFeAsO our DOS is in good agreement with previous reports[24, 30, 48].

It is well recognized that electrons near the Fermi surface contribute to the superconductivity. From the analysis of Fig. 2 around the Fermi level in the range -3 eV to 2 eV we found Fe 3d states while As 4p and O 2p states appear at lower energies, from -2 eV down to -5 eV. Actually As 4p states may hybridize with Fe 3d near the Fermi surface while the O 2p contribution can be neglected. In the Co doped system, $Co^{2+}$ ($3d^7$) ions contribute with one more electron respect to $Fe^{2+}$ ($3d^6$) and the Fermi level is pushed up of about 0.7 eV. We can see that the DOS of LaFeAsO near the Fermi energy is monotonous with the energy, while in the LaCoAsO the Fermi level locates near the $t_2$ peak, and the Fermi level both locate at the steep edge. We may address that the changes near the Fermi level are correlated to the spin density wave transition in the LaFeAsO near 150 K[5, 14, 49] and in the LaCoAsO at 66 K[44] which make these parent materials out from superconductivity.

Both Fe and Co 3d states split by the exchange interaction and the crystal field and, in the ideal TMAs tetragonal structure, the crystal field due to the four nearest neighbor As atoms splits the fivefold degenerate d states of a free TM atom into a doubly degenerate e band ($d_{z2}$ and $d_{x2-y2}$) and a triply degenerate $t_2$ band ($d_{xy}$, $d_{xz}$ and $d_{yz}$). Similarly, in TM-doped GaN diluted magnetic semiconductor systems, the large interaction among TM $t_2$ states and N 2p states pushes up the $t_2$ band above the e band[50-52]. However, from the orbital resolved DOS of Fe and Co 3d states in the parent compounds shown in Fig. 3, all the fivefold degenerate d states contributes around the Fermi energy indicating that the hybridization and the crystal field are relatively small in the distorted tetrahedral TMAs system. Moreover, the complex distributions of TM 3d states are influenced by the TM-TM direct interactions with bond length of 2.85 Å in the same layers. The complicate distribution of Fe or Co 3d states is different from $CuO_2$ planes. In a comparison with the copper-oxide superconductor systems, a $Cu^{2+}$ occupies a planar fourfold square site and the DOS at the Fermi level are mainly correlated with the $d_{x2-y2}$ orbital, suggesting a different superconductive mechanism between cuprate and new FeAs based superconductors.

The band structures of LaFeAsO and LaCoAsO are showed in Fig 4. The band structure of LaFeAsO is in good agreement with previous data[24, 48], and the lack of dispersion along both Γ-Z and A-M directions suggests a quasi 2-D structure, in consistent with the Bader charge analysis. The band shape of these compounds is quite similar, except at the Fermi energy where it is pushed up about 0.7 eV in the LaCoAsO, in agreement with the DOS data. To clarify the role of electrons at the Fermi surface we compare in Fig 5 the band structures of LaFeAsO and LaCoAsO in the range -0.6 eV to 0.6 eV. The dispersion near Fermi level between these two parent compounds is greatly different as a result of the Fermi level shift, which may result in different superconductivity mechanics.

**II LaCo$_x$Fe$_{1-x}$AsO system**

In the experiment reported by Wang et al.[27], Co replaces Fe in a single crystalline phase as shown by XRD. According to our lattice and internal parameter relaxation steps, we claim that the doped system is energetically stable for the partial substitution of cobalt for iron. We find that the two kinds of As-Co-As angels change to 107.41° and 113.68°, trend to less distorted tetrahedral than the parent compounds. The Bader analysis data are very close to the pure phase.

As discussed for the parent materials, the Fermi level will push up about 0.7 eV for the replacement of Co for Fe in LaFeAsO. Thus, it is expected that Fermi level would locate at the range in the doping system. In Fig. 6 we compare the total DOS of both LaCo$_{0.125}$Fe$_{0.875}$AsO and LaCo$_{0.25}$Fe$_{0.75}$AsO and those of the pure LaFeAsO and LaCoAsO compounds. Like for other calculations on Fe-based superconductors such as LaFeAsO[24, 30, 48] and LaFePO[47], the Fermi energy lies just above a peak in the DOS. Actually, their DOS have a very steep and negative slope near the Fermi level, which drives the system close to a magnetic instability and at the moment no Fe-based parent materials exhibit superconductivity. At higher Co concentration the occupation at Fermi level decreases, i.e., 4.85 for the pure LaFeAsO, 3.25 for LaCo$_{0.125}$Fe$_{0.875}$AsO and 1.5 for LaCo$_{0.25}$Fe$_{0.75}$AsO, and the DOS near the Fermi level become flat pushing the doped system away from the magnetic instability. Many experiments in other superconductor systems revealed that the appearance of spin density wave transition may destroy superconductivity in Fe-based compounds; however nothing is known for these systems and it is fundamental to explore the relationship between SDW and superconductivity in these Fe-based compounds.

**Conclusions**

We have investigated the geometry structure, binding properties and electronic structure of LaFe$_{1-x}$Co$_x$AsO and its parent materials through first principle calculations. We find that FeAs-based ZrCuSiAs structure compounds LaFeAsO, LaCoAsO and the cobalt doping systems have quasi two-dimensional character, with ionic layer-layer interaction. The Fe or Co 3d states mainly contribute near the Fermi level, and the orbit resolved components of 3d states distribute complicate, which reveals an underlying mechanics different from cuprate superconductors. We further compare the DOS near the Fermi energy of LaFeAsO, LaCoAsO and the cobalt doping compounds, and demonstrate that cobalt doping push up the Fermi level of the LaFeAsO from a steep and negative edge towards the flat distribution of doping system, which suppresses the spin density wave transition in $LaFe_{1-x}Co_xAsO$ doping systems. It would be interesting to explore the relationship between spin fluctuations and superconductivity for future work.

**Acknowledgements**

This work is partially supported by the National Outstanding Youth Fund (Project No. 10125523 to Z.W.), by the Knowledge Innovation Program of the Chinese Academy of Sciences (KJCX2-SW-N11), and by Supercomputing Center, CNIC, CAS. We are grateful for discussions with A. Marcelli.

Table 1. Comparison of experimental and calculated crystal lattice and internal parameters of LaFeAsO and LaCoAsO compounds.

|  |  | $a$ (Å) | $c$ (Å) | $z_{La}$ | $z_{As}$ |
|---|---|---|---|---|---|
| LaFeAsO | Calculated | 4.026 | 8.611 | 0.1451 | 0.6381 |
|  | Experiment[5] | 4.035 | 8.741 | 0.1415 | 0.6512 |
|  | Experiment [13] | 4.038 | 8.754 | - | - |
|  | Experiment [14] | 4.030 | 8.737 | 0.1418 | 0.6507 |
| LaCoAsO | Calculated | 4.045 | 8.518 | 0.1454 | 0.6382 |
|  | Experiment [13] | 4.054 | 8.472 | - | - |

Table 2. Comparison of the electronic charges to different species obtained by the Bader analysis and their pure ionic picture.

|  | La | TM | As | O | LaO | TMAs |
|---|---|---|---|---|---|---|
| Bader (LaFeAsO) | 9.1009 | 7.7103 | 5.8928 | 7.2960 | 16.3969 | 13.6031 |
| Ionic picture (LaFeAsO) | 8($La^{3+}$) | 6($Fe^{2+}$) | 8($As^{3-}$) | 8($O^{2-}$) | 16 | 14 |
| Bader (LaCoAsO) | 9.0957 | 8.8937 | 5.7158 | 7.2947 | 16.3904 | 14.6095 |
| Ionic picture (LaCoAsO) | 8($La^{3+}$) | 7($Co^{2+}$) | 8($As^{3-}$) | 8($O^{2-}$) | 16 | 15 |

Figure 1. Cystal structure of the LaFe$_{1-x}$Co$_x$AsO. Elements are labeled inside the spheres.

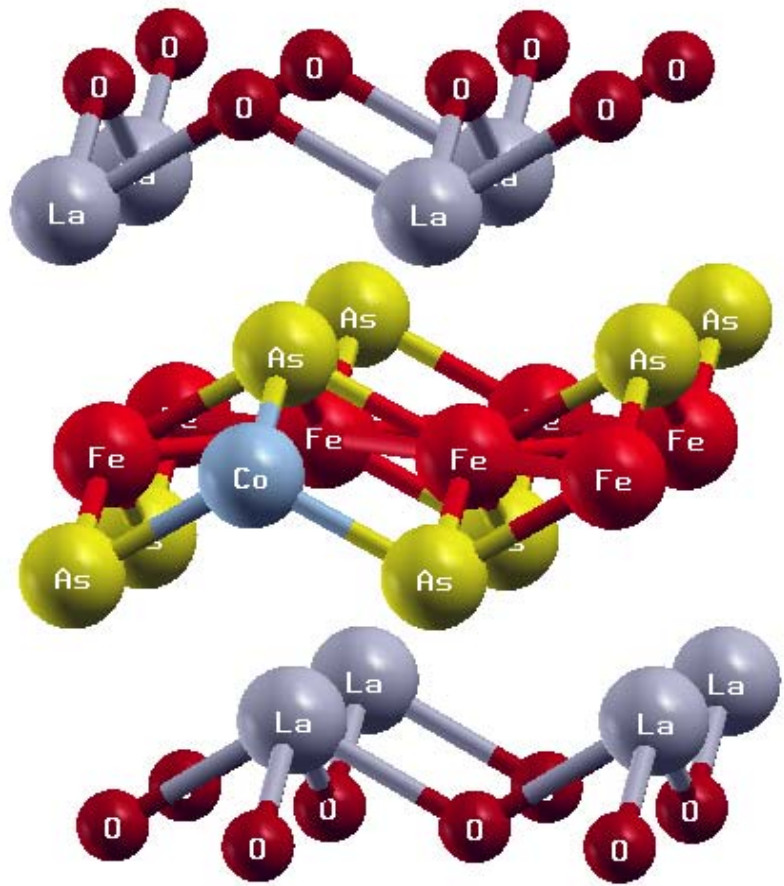

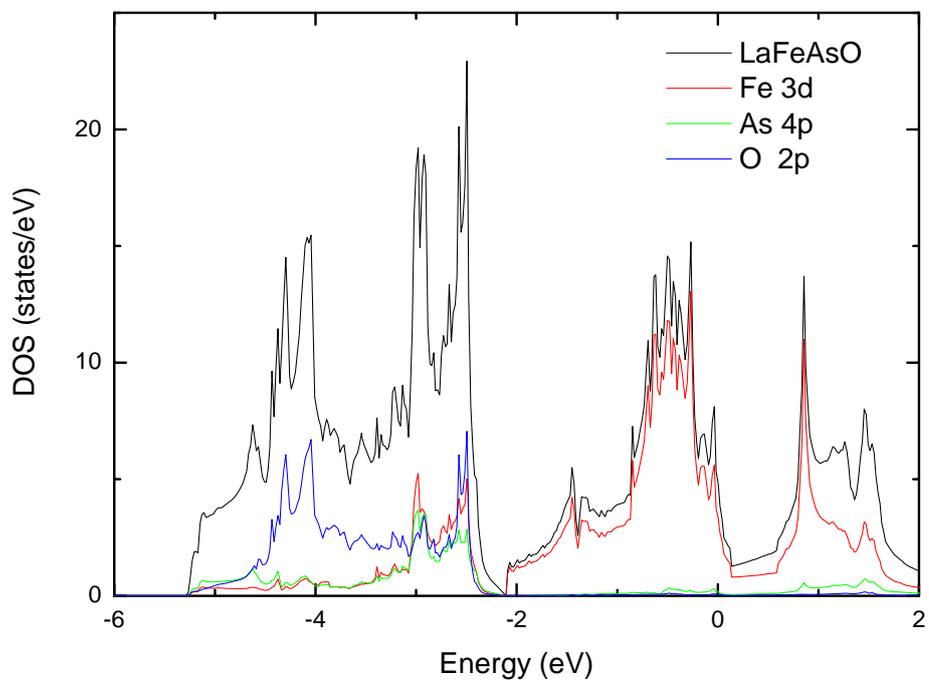

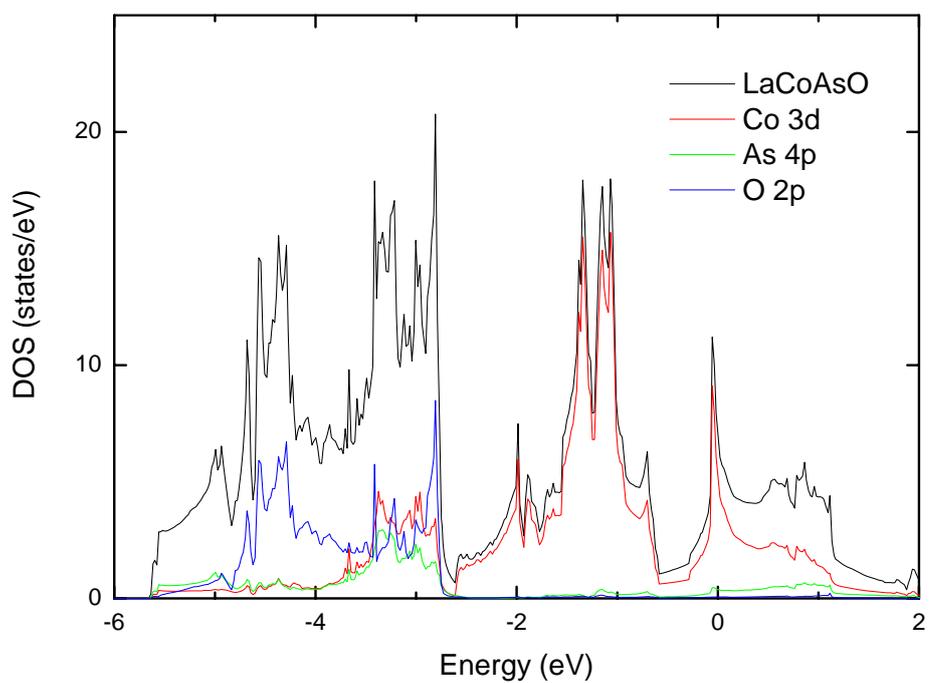

Figure 2. Total density of states of LaFeAsO (top) and LaCoAsO (bottom) and their partial density of states within the GGA approximation. The character of Fe (Co), O and As is shown separately. All energies are relative to the Fermi energy.

Figure 3. Orbital resolved DOS of Fe 3d states in the LaFeAsO (top) and of Co 3d states in LaCoAsO (bottom).

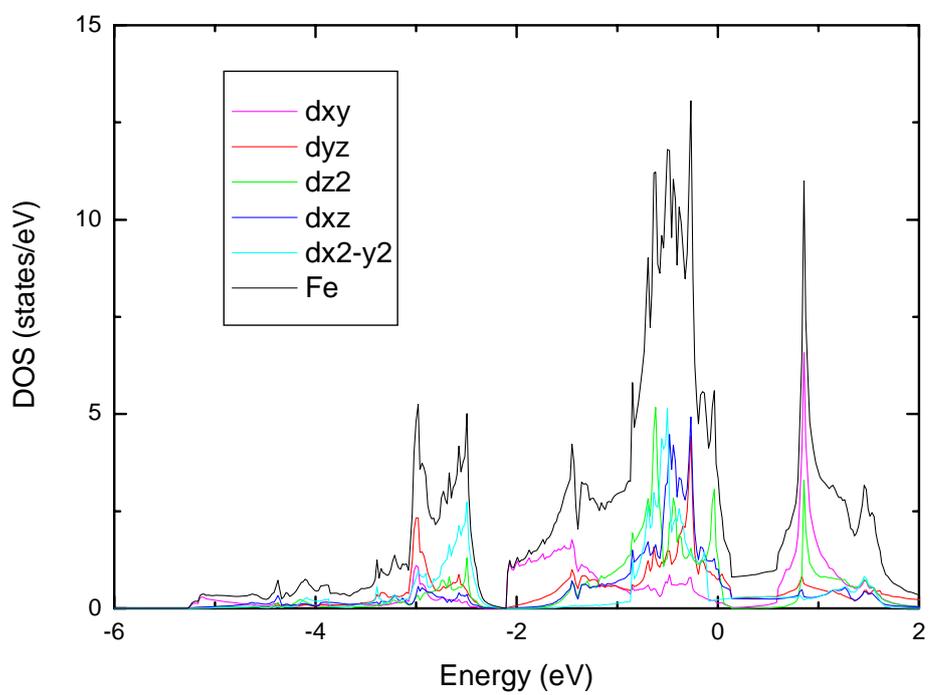

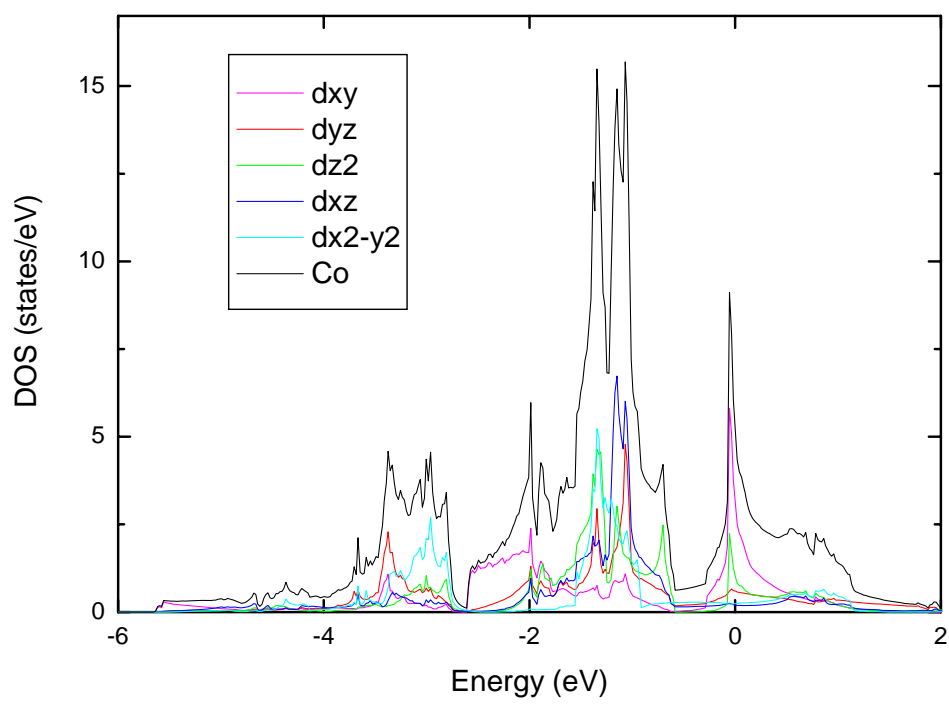

Figure 4. Calculated band structures of LaFeAsO (top) and LaCoAsO (bottom). All energies are relative to the Fermi energy.

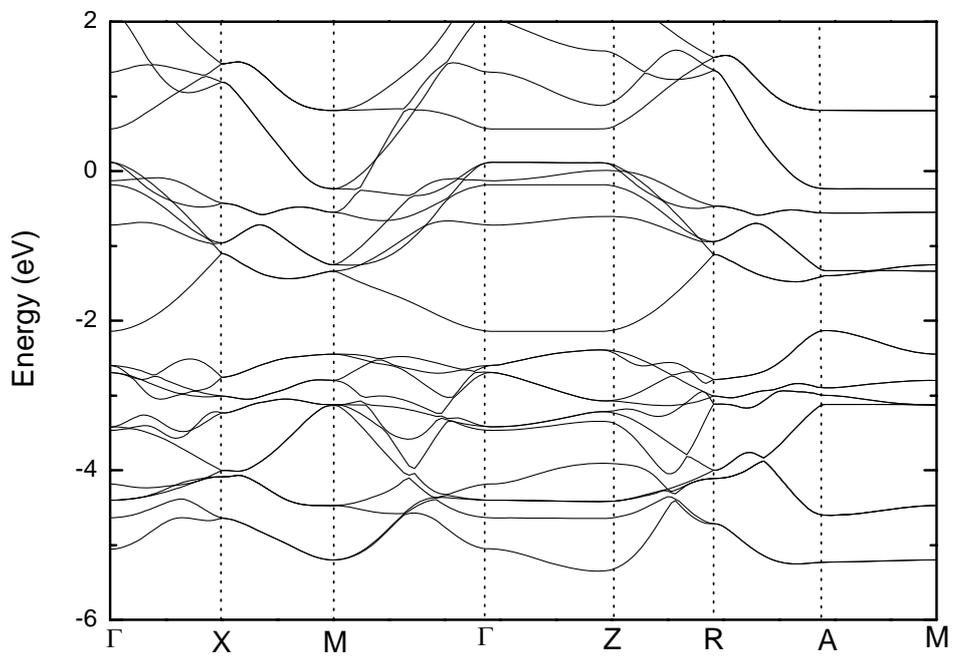

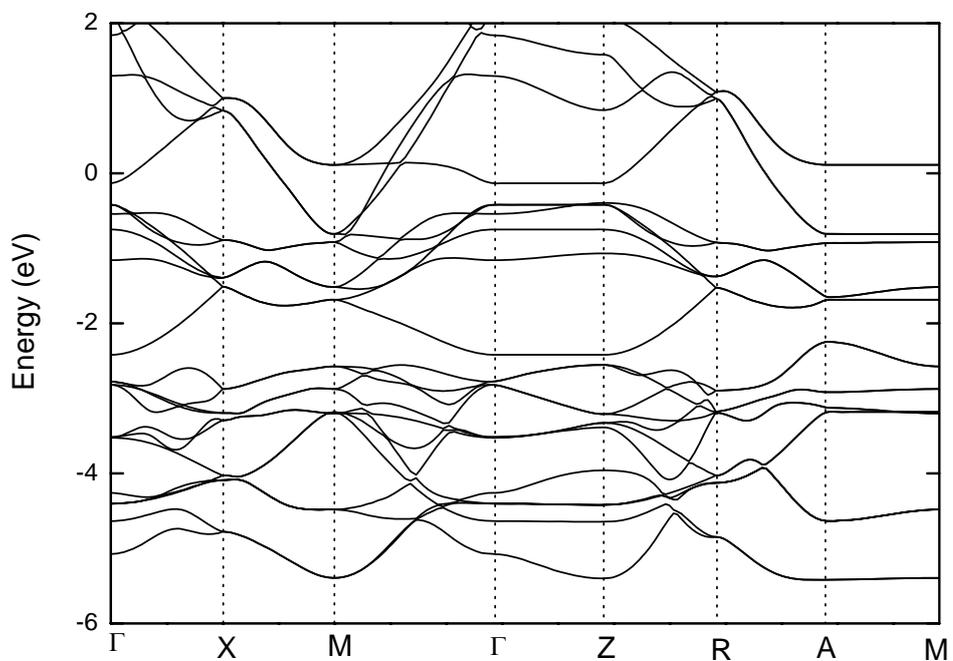

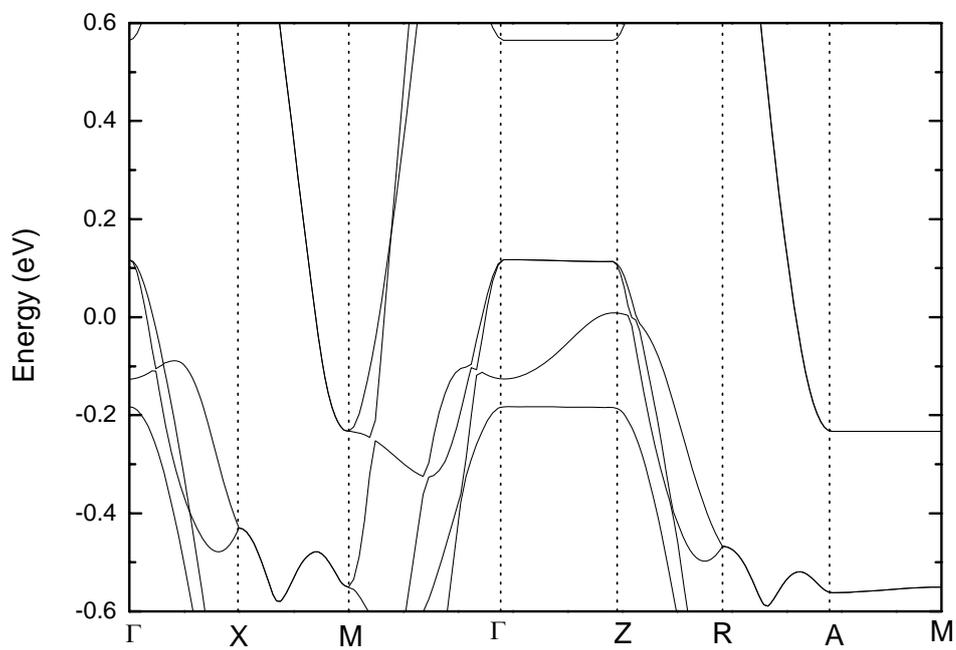

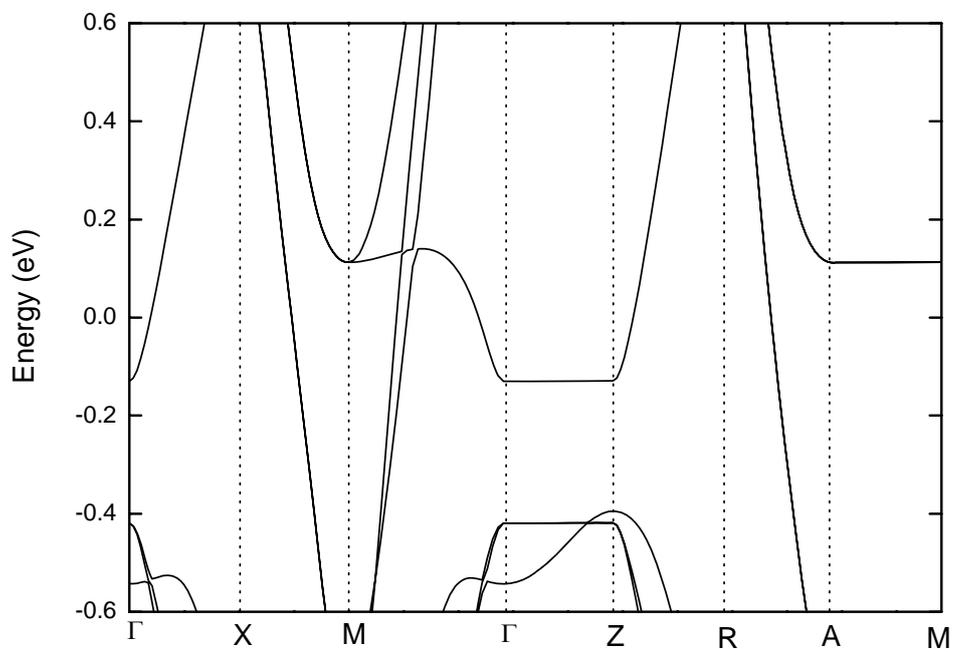

Figure 5. Band structures of LaFeAsO (top) and LaCoAsO (bottom) near the Fermi energy from -0.6 eV to +0.6 eV. All energies are relative to the Fermi level.

Figure 6. Comparison of the total density of states of LaCo$_x$Fe$_{1-x}$AsO, LaFeAsO and LaCoAsO. All energies are relative to the Fermi level.

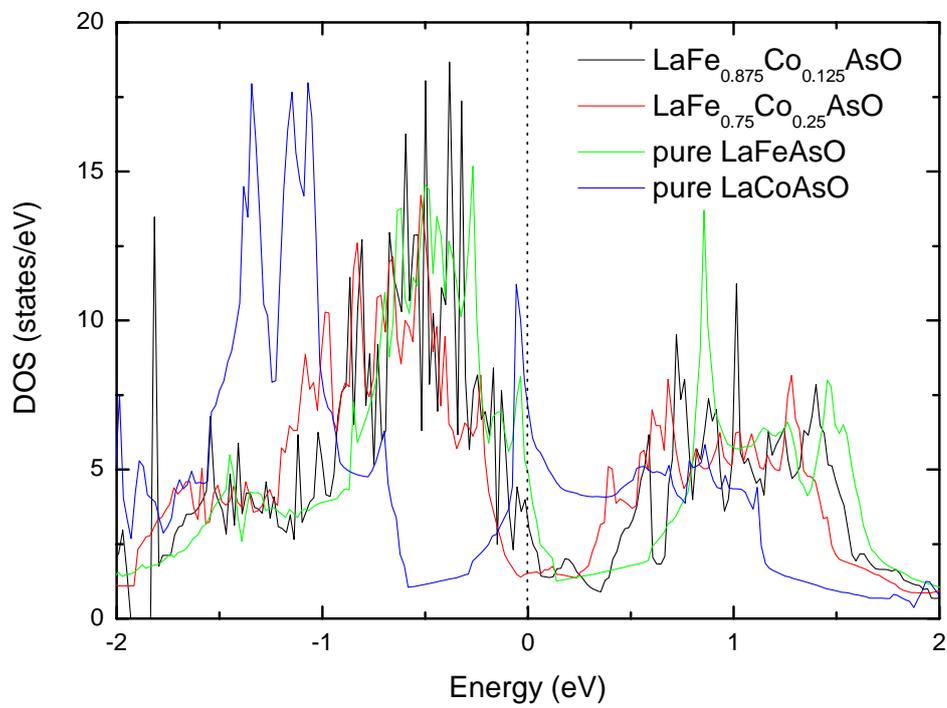